\documentclass[12pt]{article}
\usepackage[totalwidth=470truept,totalheight=636truept]{geometry}
\usepackage{latexsym,graphicx,amsfonts,amssymb,amsmath,color,dsfont,cancel,tikz,multirow,slashed}

\usepackage[format=hang,font=small,textfont=it]{caption}

\usepackage[hypertexnames=false,hidelinks]{hyperref}

\DeclareMathAlphabet{\mymathbb}{U}{BOONDOX-ds}{m}{n}

\newcommand{\sect}[1]{\setcounter{equation}{0}\section{#1}}

\global\arraycolsep=1truept

\hypersetup{colorlinks=true,
    linkcolor=black,
    filecolor=black, 
    citecolor=black,
    urlcolor=blue}

\RequirePackage{booktabs}

\begin{document}

\null

\vskip1truecm

\begin{center}
{\Large \textbf{Renormalization Of Massive Rank-1 Field Theory}}

\vskip.8truecm

{\Large \textbf{Nonminimally Coupled To Quantum Gravity}}

\vskip1truecm

\textsl{{\large Marco Piva}}

\vskip .1truecm

{\textit{Faculty of Physics, University of Warsaw,\\ Pasteura 5, 02-093 Warsaw, Poland}}

\vspace{0.2cm}

mpiva@fuw.edu.pl
\vskip1truecm

\textbf{Abstract}
\end{center}

We consider a self-interacting, massive rank-1 field coupled to quantum gravity. 
The theory  is renormalizable by power counting and contains a massive spin-1 field and a massive scalar field. The latter has a propagator with negative residue and it is quantized as a purely virtual particle, namely it cannot appear as external on-shell state but contributes to renormalization. In this way the resulting theory is also unitary, both in flat spacetime and when it is coupled to renormalizable quantum gravity (where purely virtual particles are required as well). We compute the full set of one-loop beta functions, including those of the nonminimal couplings. We show that the gravitational couplings cannot be made asymptotically free in absence of tachyons, even with the addition of the two nonminimal terms. Various aspects of renormalizability are discussed by studying the beta functions. Finally, we compare our model with Proca theory, where renormalizability is spoiled if self interactions are present.

\vfill\eject

\sect{Introduction}
Abelian massive vector fields have a variety of phenomenological applications that span from inflation~\cite{Golovnev:2008cf,Golovnev:2008hv,Bertolami:2015wir,Oliveros:2016myr} to dark matter~\cite{Graham:2015rva,Ahmed:2020fhc,Barman:2021qds}. The typical adopted model is Proca theory~\cite{Proca:1936fbw}, which can be shown to be renormalizable in flat spacetime by introducing the so-called \emph{Stueckelberg field}~\cite{Stueckelberg:1938hvi}. However, this holds only if the interaction terms have at most dimension four and are invariant under an Abelian $U(1)$ symmetry~\cite{Ruegg:2003ps}. Therefore, if we introduce self interactions, renormalizability is spoiled. Moreover, the coupling to (renormalizable) quantum gravity generates all the nonminimal couplings that are allowed by power counting, even if they are absent at tree level. 

On the other hand, a theory of general massive vectors that transform as the irreducible representation $\left(\frac{1}{2},\frac{1}{2}\right)$ of the Lorentz group (which we call \emph{rank-1 field}) is renormalizable by power counting, even when self interactions are present, as well as if it is coupled to higher-derivative quantum gravity. In this theory, all the four degrees of freedom of the field propagate and are decomposed into a massive spin-1 field and a massive scalar field. The latter, if quantized with the usual Feynman prescription, is a ghost and leads to a non-unitary $S$ matrix. However, unitarity can be restored by adopting the so-called \emph{fakeon prescription}~\cite{Anselmi:2017yux, Anselmi:2017ygm}, which turns the ghost into a \emph{purely virtual particle}, that is to say, a particle which cannot appear as external on-shell state at any energy scale but can circulate inside Feynman diagrams. Moreover, the fakeon prescription allows us to keep the renormalizability properties untouched, since the Euclidean sector of the theory is not changed.

The theory discussed in this paper belongs to a more general class recently studied in~\cite{Anselmi:2020opi} and~\cite{Piva:2021nyj}, where irreducible multiplets of arbitrary spin are considered. In particular, in~\cite{Piva:2021nyj} the models were coupled to renormalizable quantum gravity and their contribution to the gravitational beta functions derived. However, certain simplifications were adopted. For example, nonminmal couplings to gravity have been neglected. 

In this paper we study the full renormalization of the rank-1 field theory, including all the nonminimal couplings and a quartic self-interaction term. We derive the beta functions of all parameters and discuss their limits. The computations are performed by means of Feynman diagrams and with the help of the Batalin-Vilkovisky formalism~\cite{Batalin:1981jr, Batalin:1983ggl} to handle nonmultiplicative renormalization. We also compute counterterms in the case of Proca theory to show that it is not renormalizable.

We stress that most of the literature on this topic consider Proca theory, or one of its self-interacting version~\cite{BeltranJimenez:2016rff}, coupled to external gravity~\cite{Toms:2015fja, Buchbinder:2017zaa,Ruf:2018vzq,Garcia-Recio:2019iia}. In this sense our work is more complete since it accounts for the effects of quantum gravity. Clearly, those contributions depends on the chosen theory of quantum gravity.

Besides possible phenomenological applications, the rank-1 field theory can be used as prototype for the study of Stelle theory~\cite{Stelle:1976gc} and quantum gravity with fakeons~\cite{Anselmi:2018tmf} in the context of scattering amplitudes. In fact, it mimics those theories in several aspects, although there are no higher derivatives. For example, in addition to the crucial cancellations between the massive spin-1 and the massive scalar ghost/fakeon, the 2-to-2 tree-level scattering amplitude grows as a power of the center-of-mass energy squared, pretty much like in the case of Stelle gravity~\cite{Dona:2015tra}. However, we do not address this matter and concentrate only on the properties under renormalization, which hold regardless of the fakeon prescription.

The paper is organized as follows. In~\autoref{sec:rank1} we recall the main properties of rank-1 field theory and discuss its renormalizability in flat spacetime within two different formulations, one of them including a field for the massive scalar. In~\autoref{sec:Stue} we highlight the differences with the Proca model in the Stueckelberg approach. In~\autoref{sec:BV} we review the quantization of higher-derivative quantum gravity in the Batalin-Vilkovisky formalism. In~\autoref{sec:couplingQG} we couple the self-interacting rank-1 field to quantum gravity and derive the one-loop counterterms and beta functions for all the parameters. Moreover, we derive the counterterms in the case of Proca theory and show that higher-dimensional operators are generated at one loop. For completeness, we also report all the beta functions in the case of a nonminimally coupled, self-interacting scalar. Finally,~\autoref{sec:concl} contains our conclusions.  

\emph{Notation and conventions:} We use the signature $(+,-,-,-)$ for the metric tensor. The Riemann and Ricci tensors are defined as $R^{\mu}_{ \ \nu\rho\sigma}=\partial_{\rho}\Gamma^{\mu}_{\nu\sigma}-\partial_{\sigma}\Gamma^{\mu}_{\nu\rho}+\Gamma^{\mu}_{\alpha\rho}\Gamma^{\alpha}_{\nu\sigma}-\Gamma^{\mu}_{\alpha\sigma}\Gamma^{\alpha}_{\nu\rho}$ and $R_{\mu\nu}=R^{\rho}_{ \ \mu\rho\nu}$, respectively. We write the four-dimensional integrals over spacetime points of a function $F$ of a field $\phi(x)$ as $\int\sqrt{-g}F(\phi)\equiv\int\text{d}^4x\sqrt{-g(x)} F\left(\phi(x)\right)$. We always assume that the integral of the Gauss-Bonnet term vanish, i.e. $\int\sqrt{-g}\left(R_{\mu\nu\rho\sigma}R^{\mu\nu\rho\sigma}-4R_{\mu\nu}R^{\mu\nu}+R^2\right)=0.$

\sect{Massive rank-1 field theory}
\label{sec:rank1}
We consider a field $A_{\mu}$ that transforms as the irreducible representation of the Lorentz group $(\frac{1}{2},\frac{1}{2})$. The most general quadratic action for such a field is 
\begin{equation}\label{eq:rank1flatF}
    S_1(A)=\int\left(-\frac{1}{4}F_{\mu\nu}F^{\mu\nu}-\frac{\lambda_2}{2}\partial^{\mu}A_{\mu}\partial^{\nu}A_{\nu}+\frac{1}{2}m^2A_{\mu}A^{\mu}\right),
\end{equation}
where $\lambda_2$ and $m$ are real parameters. In momentum space, the propagator reads
\begin{equation}\label{eq:propA}
    D_{\mu\nu}(p)=\frac{-i}{p^2-m^2+i\epsilon}\left(\eta_{\mu\nu}-\frac{p_{\mu}p_{\nu}}{m^2}\right)+\left.\frac{-i}{p^2-\frac{m^2}{\lambda_2}}\right|_{\text{f}}\frac{p_{\mu}p_{\nu}}{m^2},
\end{equation}
where the subscript ``f" denotes that we chose the fakeon prescription for the pole associated to that term. Since most of this paper deals with renormalization, which is unchanged by the fakeon prescription, we refer the reader to the review~\cite{Piva:2023bcf} for details. What is important to know is that the fakeon prescription is a way of deforming the $S$ matrix so that it is unitary in the Fock subspace where the fakeons are removed from being external lines of Feynman diagrams, provided that all the potential ghosts are treated this way and tachyons are absent. 

 We study the residues of the propagator~\eqref{eq:propA} at the poles $p^2=m^2$ and $p^2=m_{0}^2\equiv m^2/\lambda_2$. In order to determine them we choose the rest frame $p_{\mu}=\left(m,0,0,0\right)$ and write the propagator as a $4\times4$ matrix $\mathbb{D}$. Then the residue matrices read
\begin{equation}
    -i(p^2-m^2)\mathbb{D}\left.\right|_{p^2=m^2}=\left(\setlength\arraycolsep{5pt}\begin{array}{cccc}
0 & 0 & 0 & 0\\
0 & 1 & 0 & 0\\
0 & 0 & 1 & 0\\
0 & 0 & 0 & 1\\
\end{array}\right), \quad
  -i(p^2-m_0^2)\mathbb{D}\left.\right|_{p^2=m_0^2}=\left(\setlength\arraycolsep{5pt}\begin{array}{cccc}
\frac{-1}{\lambda_2} & 0 & 0 & 0\\
0 & 0 & 0 & 0\\
0 & 0 &0 & 0\\
0 & 0 & 0 & 0\\
\end{array}\right).
\end{equation}
From the number of independent residues we can see that the first pole is associated to a massive vector, while the second one is associated to a massive scalar. Moreover, the sign of the residue of the scalar depends on $\lambda_2$ and, if we want a positive residue, we need to choose $\lambda_2<0$. However, in that case the mass squared would be negative. Since the fakeon prescription cannot be applied to tachyons, the only possibility to preserve unitarity without spoiling renormalizability is to fix $\lambda_2>0$ and then make the scalar ghost purely virtual with the fakeon prescription. 

In the standard approach, such ghost is removed by fixing $\lambda_2=0$ so $m_0$ is sent to infinity and~\eqref{eq:rank1flatF} reduces to Proca theory, which is renormalizable if the interactions are $U(1)$ invariant. However, our aim is to preserve renormalizability and unitarity when self interactions, as well as nonminimal couplings to gravity, are included, which cannot be done in Proca theory.

Finally, we show that it is possible to write an action, equivalent to~\eqref{eq:rank1flatF}, where the scalar field is explicit. First, we introduce an auxiliary field $\varphi$ and define a new action
\begin{equation}\label{eq:rank1flataux}
    S_1^{\prime}(A,\varphi)=\int\left[-\frac{1}{4}F_{\mu\nu}F^{\mu\nu}+\frac{1}{2}m^2A_{\mu}A^{\mu}+\frac{m^2}{2\lambda_2}\left(\varphi^2-2\frac{\lambda_2}{m}\partial^{\mu}A_{\mu}\varphi\right)\right].
\end{equation}
The equivalence with~\eqref{eq:rank1flatF} is obtained once the equations of motion for $\varphi$ are used, i.e.
\begin{equation}
    S_1^{\prime}(A,\varphi(A))=S_1(A), \qquad \varphi(A)=\frac{\lambda_2}{m}\partial^{\mu}A_{\mu}.
\end{equation}
Then we apply the redefinition
\begin{equation}\label{eq:StueRed}
    A_{\mu}\rightarrow A_{\mu}-\frac{1}{m}\partial_{\mu}\varphi
\end{equation} 
and obtain
\begin{equation}\label{eq:rank1flatfakeon}
S_1^{\prime}(A,\varphi)=\int\left(-\frac{1}{4}F_{\mu\nu}F^{\mu\nu}+\frac{1}{2}m^2A_{\mu}A^{\mu}-\frac{1}{2}\partial_{\mu}\varphi\partial_{\mu}\varphi+\frac{1}{2}\frac{m^2}{\lambda_2}\varphi^2\right),
\end{equation}
where the negative sign of the kinetic term for the field $\varphi$ is now manifest. If we include self interactions in~\eqref{eq:rank1flatF} renormalizability is not spoiled even in the variables~\eqref{eq:rank1flatfakeon} because of the presence of the ghost/fakeon, whose negative residue is crucial (see below).

Note that the limit $\lambda_2=0$ of~\eqref{eq:rank1flatF} gives Proca theory, which in~\eqref{eq:rank1flatfakeon} is obtianed by setting both $\lambda_2$ and $\varphi$ to zero with $\varphi/\lambda_2$ constant, since $\varphi(A)\propto\lambda_2$. 

Before moving to the coupling to quantum gravity, it is instructive to show how renormalization works when switching between the variables~\eqref{eq:rank1flatF} and~\eqref{eq:rank1flatfakeon}. Here and in the rest of the paper, we use the dimensional regularization and define $\varepsilon=4-D$, where $D$ is the continued spacetime dimension. For simplicity we consider only a quartic self interaction in flat spacetime, so the total action reads 
\begin{equation}\label{eq:rank1FlatFInt}
      S_1(A)=\int\left[-\frac{1}{4}F_{\mu\nu}F^{\mu\nu}-\frac{\lambda_2}{2}\partial^{\mu}A_{\mu}\partial^{\nu}A_{\nu}+\frac{1}{2}m^2A_{\mu}A^{\mu}+\frac{\lambda_4}{8}(A_{\mu}A^{\mu})^2\right].
\end{equation}
At one loop the renormalized action is obtained by means of the redefinitions 
\begin{equation}
    m^2\rightarrow Zm^2, \qquad \lambda_4\rightarrow Z_4\lambda_4,
\end{equation}
with
\begin{equation}
    Z=1-\frac{3\lambda_4(1+3\lambda_2^2)}{32\pi^2 \lambda_2^2\varepsilon}+\mathcal{O}(\lambda_4^2), \qquad Z_4=1-\frac{\lambda_4(5+2\lambda_2+17\lambda_2^2)}{32\pi^2\lambda_2^2\varepsilon}+\mathcal{O}(\lambda_4^2).
\end{equation}
The field $A_{\mu}$ and the parameter $\lambda_2$ are not renormalized at this order, since the one-loop two-point function is given only by a tadpole diagram, which is proportional to the mass.

Following the procedure used to obtain~\eqref{eq:rank1flatfakeon}, in the interacting case we obtain the action
\begin{equation}\label{eq:rank1flatfakeonInt}
S_1^{\prime}(A,\varphi)=\int\left[-\frac{1}{4}F_{\mu\nu}F^{\mu\nu}+\frac{1}{2}m^2A_{\mu}A^{\mu}-\frac{1}{2}\partial_{\mu}\varphi\partial^{\mu}\varphi+\frac{1}{2}\frac{m^2}{\lambda_2}\varphi^2+\frac{\lambda_4}{8}(A_{\mu}A^{\mu})^2\left.\right|_{A\rightarrow A-\frac{\partial\varphi}{m}}\right].
\end{equation}
The shift in the last term introduces several interactions of dimension greater than four. However, as mentioned above, those interactions do not spoil renormalizability. This is due to the fact that the $\varphi$ propagator has a negative residue, which allows for crucial cancellations between diagrams. In the original variables~\eqref{eq:rank1FlatFInt} those cancellations are evident from the propagator~\eqref{eq:propA}, which behaves like $1/p^2$ in the ultraviolet (UV). 

To give an explicit example, we compute the one-loop counterterms and show how the renormalization constants relate to each other in the two cases.

The general expression for the renormalized action is
\begin{equation}\label{eq:S1reno}
\begin{split}
S_{1,\text{R}}^{\prime}(A,\varphi)=\int \Big\{&-\frac{Z_{20}}{4}F_{\mu\nu}F^{\mu\nu}+\frac{Z_{20}^{\prime}}{2}m^2A_{\mu}A^{\mu}-\frac{Z_{02}}{2}\partial_{\mu}\varphi\partial^{\mu}\varphi+\frac{Z_{02}^{\prime}}{2}\frac{m^2}{\lambda_2}\varphi^2+mZ_{11}A_{\mu}\partial^{\mu}\varphi\\
 &+\frac{\lambda_4}{8}\Big[Z_{40}(A_{\mu}A^{\mu})^2-\frac{4Z_{31}}{m}A_{\mu}\partial^{\mu}\varphi A_{\nu}A^{\nu}+\frac{6Z_{22}}{m^2}\partial_{\mu}\varphi\partial^{\mu}\varphi A_{\nu}A^{\nu}\\
 &-\frac{4Z_{13}}{m^3}\partial_{\mu}\varphi\partial^{\mu}\varphi \partial_{\nu}\varphi A^{\nu}+\frac{Z_{04}}{m^4}(\partial_{\mu}\varphi\partial^{\mu}\varphi)^2 \Big] \Big\},
\end{split}
\end{equation}
where the subscripts in $Z$ and $Z'$ indicate the number of fields $A$ and the number of fields $\varphi$, respectively. We find 
\begin{equation}
    Z_{20}=Z_{02}^{\prime}=1, \qquad Z^{\prime}_{20}=Z_{02}=Z, \qquad Z_{11}=1-Z
\end{equation}
\begin{equation}
    Z_{40}=Z_{31}=Z_{22}=Z_{13}=Z_{04}= Z_4
\end{equation}
and the action~\eqref{eq:S1reno} reads
\begin{equation}\label{eq:S1reno2}
\begin{split}
S_{1,\text{R}}^{\prime}(A,\varphi)=\int \Big[&-\frac{1}{4}F_{\mu\nu}F^{\mu\nu}+\frac{Z}{2}m^2A_{\mu}A^{\mu}-\frac{Z}{2}\partial_{\mu}\varphi\partial^{\mu}\varphi+\frac{1}{2}\frac{m^2}{\lambda_2}\varphi^2+m(1-Z)A_{\mu}\partial^{\mu}\varphi\\
 &+\frac{\lambda_4Z_4}{8}(A_{\mu}A^{\mu})^2\left.\right|_{A\rightarrow A-\frac{\partial\varphi}{m}} \Big],
\end{split}
\end{equation}
Note that a mixing kinetic term is generated at one loop. This means that $A_{\mu}$ and $\varphi$ are renormalized by the field redefinitions
\begin{equation}
    A_{\mu}\rightarrow A_{\mu}-\frac{Z-1}{m Z}\partial_{\mu}\varphi, \qquad \varphi\rightarrow \frac{1}{\sqrt{Z}}\varphi
\end{equation}
together with
\begin{equation}
    m^2\rightarrow Zm^2, \qquad \lambda_{4}\rightarrow Z_4\lambda_4
\end{equation}
Alternatively, we can diagonalize the quadratic part of the action~\eqref{eq:S1reno2} by means of the transformation
\begin{equation}
    A_{\mu}\rightarrow A_{\mu}+\frac{Z-1}{m}\partial_{\mu}\varphi, \qquad \varphi\rightarrow Z\varphi,
\end{equation}
which leaves $A_{\mu}-\frac{1}{m}\partial_{\mu}\varphi$ invariant, and the action~\eqref{eq:S1reno2} becomes
\begin{equation}
    S_{1,\text{R}}^{\prime}(A,\varphi)=\int \Big[-\frac{1}{4}F_{\mu\nu}F^{\mu\nu}+\frac{Z}{2}m^2A_{\mu}A^{\mu}-\frac{Z}{2}\partial_{\mu}\varphi\partial^{\mu}\varphi+\frac{Z^2}{2}\frac{m^2}{\lambda_2}\varphi^2+\frac{\lambda_4Z_4}{8}(A_{\mu}A^{\mu})^2\left.\right|_{A\rightarrow A-\frac{\partial\varphi}{m}} \Big].
\end{equation}
In this variables it is clear that renormalization is multiplicative, i.e.
\begin{equation}
    m^2\rightarrow Zm^2, \qquad \lambda_4\rightarrow Z_4\lambda_4, \qquad \varphi\rightarrow \sqrt{Z}\varphi,
\end{equation}
and such that $A_{\mu}-\frac{1}{m}\partial_{\mu}\varphi$ does not renormalize (at one loop). This is in agreement with the results obtained in the variables~\eqref{eq:rank1flatF}, where the field $A_{\mu}$ does not renormalize at this order.

\sect{Proca and Stueckelberg theories}\label{sec:Stue}t
In this section we briefly review the Stueckelberg mechanism and explain why it does not work for a self-interacting Proca theory. In this way we clarify the differences between the theory considered in this paper and those explored in the literature~\cite{Toms:2015fja, BeltranJimenez:2016rff, Buchbinder:2017zaa,Ruf:2018vzq,Garcia-Recio:2019iia}.

The free Proca action is
\begin{equation}
    \label{eq:procaact}
     S_{\text{P}}(A)=\int\left(-\frac{1}{4}F_{\mu\nu}F^{\mu\nu}+\frac{1}{2}m^2A_{\mu}A^{\mu}\right)
\end{equation}
and its propagator is given by the first term of~\eqref{eq:propA}, which goes like $1/m^2$ in the UV, apparently breaking power-counting renormalizability. However, this depends on the symmetry of the interactions. We can perform the redefinition~\eqref{eq:StueRed} to the free Proca action and obtain the Stueckelberg action
\begin{equation}\label{eq:StuAct}
    S_{\text{S}}(A,\varphi)=\int\left(-\frac{1}{4}F_{\mu\nu}F^{\mu\nu}+\frac{1}{2}m^2A_{\mu}A^{\mu}+\frac{1}{2}\partial_{\mu}\varphi\partial^{\mu}\varphi-mA_{\mu}\partial^{\mu}\varphi\right).
\end{equation}
The new action is invariant under the gauge symmetry
\begin{equation}\label{eq:StueTrans}
    A_{\mu}\rightarrow A_{\mu}+\partial_{\mu}f, \qquad \varphi\rightarrow \varphi+mf,
\end{equation}
where $f$ is an arbitrary function. This symmetry allows us to write Proca theory as a gauge-fixed version of the theory~\eqref{eq:StuAct}. The advantage of the action~\eqref{eq:StuAct} is that renormalizability can be explicitly proved once interactions are switched on, provided that they do not break~\eqref{eq:StueTrans}.

In order to derive the propagator we first extend~\eqref{eq:StueTrans} to the Becchi-Rouet-Stora-Tyutin (BRST) symmetry 
\begin{equation}\label{eq:brststue}
    sA_{\mu}=\partial_{\mu}C, \qquad s\varphi=mC, \qquad sC=0, \qquad s\bar{C}=0, \qquad sB=0, 
\end{equation}
where $C$, $\bar{C}$ and $B$ are the ghost, antighost and Nakanishi-Lautrup fields respectively, and $s$ is the nilpotent BRST operator. Then the gauge-fixed action is
\begin{equation}
    S_{\text{gf}}=S_{\text{S}}(A,\varphi)+s\Psi,
\end{equation}
where $\Psi$ is the gauge-fixing functional
\begin{equation}
    \Psi=\int\bar{C}\left[\mathcal{G}(A,\varphi)+\alpha B\right], \qquad \mathcal{G}(A,\varphi)=\beta m \varphi+\gamma\partial^{\mu}A_{\mu}
\end{equation}
and $\alpha$, $\beta$, $\gamma$ are gauge-fixing parameters. Integrating out the field $B$, the gauge fixed action becomes
\begin{equation}
    S_{\text{gf}}=S_{\text{S}}-\frac{1}{4\alpha}\int\mathcal{G}^2+\int\bar{C}\left(\gamma\square+\beta m^2\right)C.
\end{equation}

The propagators read
\begin{equation}
    \langle A_{\mu}(p)A_{\nu}(-p)\rangle_0=\frac{-i}{p^2-m^2+i\epsilon}\left[\eta_{\mu\nu}-\frac{(\gamma^2-2\alpha)p^2+(2 \alpha +\beta^2-2\beta\gamma)m^2}{(\gamma p^2-\beta m^2)^2}p_{\mu}p_{\nu}\right],
\end{equation}

\begin{equation}
    \langle A_{\mu}(p)\varphi(-p)\rangle_0=\frac{m(2\alpha-\beta\gamma)}{\gamma p^2-\beta m^2}p_{\mu}=-\langle \varphi(p) A_{\mu}(-p)\rangle_0, \qquad \langle \varphi(p)\varphi(-p)\rangle_0=\frac{\gamma^2p^2-2\alpha m^2}{(\gamma p^2-\beta m^2)^2}.
\end{equation}
We show that by choosing different values of $\alpha$, $\beta$ and $\gamma$ we can interpolate between Proca propagator and propagators that behaves like $\sim 1/p^2$ in the UV for both  $A_{\mu}$ and $\varphi$. Indeed, by choosing $\alpha=0$ and $\gamma=0$ we find

\begin{equation}
    \langle A_{\mu}(p)A_{\nu}(-p)\rangle_0=\frac{-i}{p^2-m^2+i\epsilon}\left(\eta_{\mu\nu}-\frac{p_{\mu}p_{\nu}}{m^2}\right),
\end{equation}

\begin{equation}
    \langle A_{\mu}(p)\varphi(-p)\rangle_0=\langle \varphi(p) A_{\mu}(-p)\rangle_0=\langle \varphi(p)\varphi(-p)\rangle_0=0,
\end{equation}
while for $\alpha=\beta^2/2$ and $\gamma=\beta$ we have
\begin{equation}
    \langle A_{\mu}(p)A_{\nu}(-p)\rangle_0=\frac{-i\eta_{\mu\nu}}{p^2-m^2+i\epsilon},
\end{equation}
\begin{equation}
    \langle A_{\mu}(p)\varphi(-p)\rangle_0=\langle \varphi(p) A_{\mu}(-p)\rangle_0=0,\qquad \langle \varphi(p)\varphi(-p)\rangle_0=\frac{i}{p^2-m^2+i\epsilon}.
\end{equation}
Therefore, there is a gauge choice where the theory is manifestly renormalizable, pretty much like the case of theories with a spontaneously broken symmetry. It is clear that the interactions need to be invariant under~\eqref{eq:StueTrans} in order to be able to make such a gauge choice also in the interacting case. A possibility is to couple the theory to charged fermions through the action
\begin{equation}\label{eq:fermions}
    S_{f}(\psi,\bar{\psi})=\int\bar{\psi}\left(i\slashed{D}-m_{f}\right)\psi, \qquad \slashed{D}=\gamma^{\mu}D_{\mu}=\gamma^{\mu}\left(\partial_{\mu}+iQA_{\mu}\right),
\end{equation}
where $Q$ and $m_f$ are the charge and the mass of the fermions, respectively, and the BRST transformations~\eqref{eq:brststue} are extended with
\begin{equation}
    s\psi=iQC\psi, \qquad s\bar{\psi}=-iQ\bar{\psi}C.
\end{equation}
In this case, the action~\eqref{eq:fermions} is invariant under a $U(1)$ symmetry so it is also invariant under~\eqref{eq:StueTrans} and the theory is renormalizable. Note that, because of the symmetry~\eqref{eq:StueTrans}, the field $\varphi$ can only appear in the combination $V_{\mu}\equiv A_{\mu}-\frac{1}{m}\partial_{\mu}\varphi$. Then, since $\varphi$ is decoupled, no loop diagrams with an external $\varphi$ can be built. Therefore, the mass term does not renormalize at any order.

On the other hand, if we include self interactions, renormalizability would be spoiled. For example if we add a quartic term $(A_{\mu}A^{\mu})^2$ to~\eqref{eq:StuAct} it would break gauge invariance and Proca theory could not be viewed as a gauge-fixed version of~\eqref{eq:StuAct}. 

Another possibility is to add a quartic term of the form $(V_{\mu}V^{\mu})^2$. In this case the new action is invariant, but it would allow to build counterterms with arbitrary powers of $V_{\mu}$ such as $(V_{\mu}V^{\mu})^3$ or $(V_{\mu}V^{\mu})\square(V_{\nu}V^{\nu})$. The reason is that powers of $V_{\mu}$ higher than 2 introduce powers of $1/m$ that cannot be canceled by other terms, which, instead, happens in the case of the theory~\eqref{eq:rank1flatfakeonInt}. 

In other words, renormalizability is preserved if the $U(1)$ symmetry of the massless theory is explicitly broken only by a mass term for the field $A_{\mu}$.

The procedure explained in this section can be applied also to the non-Abelian Proca theory~\cite{Boulware:1970zc, Delbourgo:1987np}. However, in that case the Stueckelberg field is coupled to the spin-1 field and renormalizability is not guaranteed.  See also~\cite{Salcedo:2022eep} for a recent investigation.

\sect{Renormalization in the Batalin-Vilkovisky formalism}
\label{sec:BV}
In this section we review the basics of the Batalin-Vilkovisky (BV) formalism, which we use to renormalize the theory of rank-1 field coupled to quantum gravity.
The BV is useful to handle the invariance under diffeomorphisms, to deal with non-multiplicative renormalization and to formulate the renormalization procedure in a systematic way. Moreover, it handles the Ward-Takahashi-Slavnonv-Taylor identities and the closure of the gauge algebra in a compact form.

We introduce the BV formalism in the context of pure gravity, so we also recall some properties of the theory. In the next section we study the cases of rank-1 fields, Proca theory and nonminimally coupled scalars, explaining additional modifications when necessary.

The action of higher-derivative quantum gravity we consider in this paper is
\begin{equation}
S_{\text{QG}}(g)=-\frac{1}{2}\int \sqrt{-g}\left[
2\Lambda+\zeta R+\frac{1}{2\alpha}C^2-\frac{1}{6\xi}R^{2}\right] ,  \label{eq:lhd}
\end{equation}%
where $\alpha $, $\xi $, $\zeta $, $\Lambda$ are positive real
constants and $C^2\equiv C_{\mu\nu\rho\sigma}C^{\mu\nu\rho\sigma}$ is the square of the Weyl tensor.

The BRST transformations associated to diffeomorphisms is defined by including Faddeev-Popov ghosts and antighosts, $C^{\rho}$, $\bar{C}^{\sigma}$ and the Nakanishi-Lautrup fields $B^{\tau}$ and read
\begin{eqnarray}
    sg_{\mu \nu}&=&-\partial_{\mu}C^{\alpha}g_{\alpha\nu}-\partial_{\nu}C^{\alpha}g_{\mu\alpha}-C^{\alpha}\partial_{\alpha}g_{\mu\nu}\nonumber\\
sC^{\rho}&=&-C^{\sigma}\partial_{\sigma}C^{\rho}\nonumber\\
s\bar{C}^{\sigma}&=&B^{\sigma}\nonumber\\
sB^{\tau}&=&0.
\end{eqnarray}
For practical reasons all the fields are collected in
\begin{equation}
\Phi^i=(g_{\mu\nu},C^{\rho},\bar{C}^{\sigma},B^{\tau}).
\end{equation}
Moreover, we introduce a row of sources
\begin{equation}
K_i=(K_g^{\mu\nu},K^{C}_{\sigma},K_{\bar{C}}^{\tau},K_B^{\tau})
\end{equation}
that are conjugated to the fields, and define the \emph{antiparentheses} of two functionals $X(\Phi,K)$ and $Y(\Phi,K)$ as
\begin{equation}
    (X,Y)\equiv\int\left(\frac{\delta_r X}{\delta \Phi^{i}}\frac{\delta_l Y}{\delta K_{i}}-\frac{\delta_r X}{\delta K_{i}}\frac{\delta_l Y}{\delta \Phi^{i}}\right),
\end{equation}
where the subscripts $r$ and $l$ denotes the right and left functional derivatives, respectively, and the integral is over the spacetime points associated with repeated
indices. With these definitions, we extend the classical action as
\begin{equation}\label{eq:extact}
S(\Phi,K)=S_{\text{QG}}+(S_K,\Psi)+S_K,
\end{equation}
where
\begin{equation}
    S_K=-\int\mathcal{R}^iK_i=\int(\partial_{\mu}C^{\alpha}g_{\alpha\nu}+\partial_{\nu}C^{\alpha}g_{\mu\alpha}+C^{\alpha}\partial_{\alpha}g_{\mu\nu})K^{\mu\nu}_g+\int C^{\sigma}\partial_{\sigma}C^{\rho}K_{\rho}^C-\int B^{\sigma}K_{\sigma}^{\bar{C}}
\end{equation}
collects the infinitesimal transformations $\mathcal{R}^i(\Phi)$ of the fields and $\Psi(\Phi)$ is the gauge-fixing functional. We choose 
\begin{equation}\label{stellepsi}
\Psi=\int\bar{C}^{\mu }(\zeta+\square/\alpha)\left( \mathcal{G}_{\mu }-\lambda B_{\mu
}\right), \qquad \mathcal{G}_{\mu}=\eta ^{\nu \rho }\partial _{\rho }g_{\mu \nu }-(\omega+1)\eta^{\nu\rho}\partial_\mu g_{\nu\rho},
\end{equation}
where $\lambda$ and $\omega$ are gauge-fixing parameters. Then the gauge-fixing term reads
\begin{equation}
    (S_K,\Psi)=\int B^{\mu }(
\zeta+\square/\alpha) \left( \mathcal{G}_{\mu }-\lambda B_{\mu
}\right)+S_{\text{gh}},  \label{skpsi}
\end{equation}%
\begin{equation}
S_{\text{gh}}=\int \bar{C}^{\mu }\partial ^{\nu
}(\zeta+\square/\alpha)\left( g_{\mu \rho }\partial _{\nu
}C^{\rho }+g_{\nu \rho }\partial _{\mu }C^{\rho }+C^{\rho }\partial _{\rho
}g_{\mu \nu }\right) .  \label{eq:ghac}
\end{equation}
The extended action~\eqref{eq:extact} satisfies the so-called \emph{master equation}
\begin{equation}\label{eq:mastereq}
(S,S)=0,
\end{equation}
which encodes the gauge invariance of $S_{\text{QG}}$ and the closure of the transformation under diffeomorphisms. After integrating $B^{\tau}$ out, the gauge-fixing term becomes
\begin{equation}
       (S_K,\Psi)=\frac{1}{4\lambda}\int \mathcal{G}^{\mu }(
\zeta+\square/\alpha)\mathcal{G}_{\mu}+S_{\text{gh}}.
\end{equation}
At this point we can invert the quadratic operator and derive the graviton propagator. We report it in the case $\lambda=1$ and $\omega=-1/2$
\begin{equation}
    \langle h_{\mu\nu}(p)h_{\rho\sigma}(-p)\rangle_0=\frac{i\alpha I_{\mu\nu\rho\sigma}}{p^2(\alpha\zeta-p^2)}+\frac{i (\xi -\alpha ) \pi _{\mu \nu } \pi _{\rho \sigma }}{6 \left(p^2-\zeta  \alpha \right) \left(p^2-\zeta  \xi \right)},
\end{equation}
where we have omitted the prescriptions and
\begin{equation}
    I_{\mu\nu\rho\sigma}=\eta_{\mu\rho}\eta_{\nu\sigma}+\eta_{\mu\sigma}\eta_{\nu\rho}-\eta_{\mu\nu}\eta_{\rho\sigma}, \qquad \pi_{\mu\nu}=\eta_{\mu\nu}+2\frac{p_{\mu}p_{\nu}}{p^2}.
\end{equation}
However, all the computations of this paper has been done for generic values of $\lambda$ and $\omega$.

The effective action $\Gamma(\Phi, K)$ is defined as the Legendre transform $\Gamma(\Phi,K)=W(J,K)-\int\Phi^iJ_i$ of the generating functional $W(J,K)$ of the connected correlation functions with respect to $J$, where $\Phi^i=\delta_rW/\delta J_i$ and
\begin{equation}
    iW(J,K)=\ln Z(J,K), \qquad Z(J,K)=\int\left[\mathrm{d}\Phi\right] \text{exp}\left(iS(\Phi,K)-i\int\Phi^lJ_l\right),
\end{equation}
$Z(J,K)$ being the generating functional of the correlation functions. Then, also the effective action satisfies the master equation
\begin{equation}\label{eq:mastereqgamma}
    (\Gamma,\Gamma)=0,
\end{equation}
which collects the Ward-Takahashi-Slavnov-Taylor identities in a compact form. It is easy to show that a consequence of equation~\eqref{eq:mastereqgamma} is that the counterterms $S_{\text{QG}}^{\text{ct}}$ satifies the equation
\begin{equation}\label{eq:cohoct}
    (S,S_{\text{QG}}^{\text{ct}})=0.
\end{equation}
Note that the operator $(S,\cdot)$ generalizes the usual BRST transformations and, when applied to the metric it reduces to the usual diffeomorphisms. Moreover, since $(S,(S,X))=0$ for any functional $X$, a general solution of~\eqref{eq:cohoct} can be written as
\begin{equation}
    S_{\text{QG}}^{\text{ct}}=G(\Phi,K)+(S,X)
\end{equation}
where $(S,G)=0$. Finally, a theorem~\cite{Anselmi:2015niw} guarantees that all the dependencies of $G$ on the unphysical fields, i.e. $C$, $\bar{C}$, $B$ and $K_i$, can be moved to $(S,X)$, so the counterterms read
\begin{equation}
    S_{\text{QG}}^{\text{ct}}=G(g)+(S,X).
\end{equation}
The functional $G$ depends only on the metric and is gauge invariant. Therefore, it is given by the integral of all possible four-dimensional covariant operators that can be built using the metric tensor, i.e
\begin{equation}
G(g)=\frac{1}{2}\int 
\sqrt{-g}\left[ 2 \Lambda\delta Z_{\Lambda}+\zeta\delta Z_{\zeta} R+\frac{\delta Z_{1/\alpha}}{2\alpha}C^2 -\frac{\delta Z_{1/\xi} }{6\xi}R^{2}%
\right],  \label{scount}
\end{equation}
where $\delta Z_i$ are real constants. Finally, the term $(S,X)$ can be removed by a \emph{canonical transformation}, i.e. a map
\begin{equation}
\Phi^i\rightarrow \Phi^{i\prime}(\Phi,K), \qquad K_i\rightarrow K^{\prime}_i(\Phi,K),
\end{equation}
such that 
\begin{equation}
    (X',Y')'=(X,Y),
\end{equation}
where the prime superscript denotes the quantities that are evaluated using the transformed fields and sources. A canonical transformation is generated by a fermionic functional $\mathcal{F}(\Phi,K^{\prime})$ such that 
\begin{equation}
    \Phi^{i\prime}=\frac{\delta \mathcal{F}}{\delta K^{\prime}_i}, \qquad K_i=\frac{\delta \mathcal{F}}{\delta \Phi^i}.
\end{equation}
In particular, any functional of the form $(S,X)$ can be removed by a canonical transformation generated by
\begin{equation}
    \mathcal{F}(\Phi,K^{\prime})=\int\Phi^iK_i^{\prime}-X(\Phi,K^{\prime}),
\end{equation}
which gives
\begin{equation}\label{eq:canonical}
    S^{\text{ct} \ \prime}_{\text{QG}}(\Phi^{\prime},K^{\prime})=S^{\text{ct}}_{\text{QG}}(\Phi,K)-(S,X(\Phi,K)).
\end{equation}
This transformation is important when we derive the counterterms by splitting the metric as $g_{\mu\nu}=\eta_{\mu\nu}+2h_{\mu\nu}$ and using Feynman diagrams. In fact, the results obtained in this way would not be covariant, in general, since they include the terms $(S,X)$ that depends on the graviton field only and are not covariant. Only after performing the canonical transformation it is possible to obtain covariant counterterms and identify the correct renormalization constants. This generalizes the more common multiplications by wave function
renormalization constants and amounts to perform the field redefinitions
\begin{equation}\label{eq:fieldred}
    g_{\mu\nu}\rightarrow g_{\mu\nu}-\Delta g_{\mu\nu}, \qquad C^{\rho}\rightarrow C^{\rho}-\Delta C^{\rho},
\end{equation}
\begin{equation}\label{eq:deltagdeltac}
   \Delta g_{\mu\nu}=t_0 g_{\mu\nu}+t_1 h_{\mu\nu}+t_2\eta_{\mu\nu}h^{\rho}_{\rho}+\mathcal{O}(h^2), \qquad \Delta C^{\rho}=s_1 C^{\rho}+\mathcal{O}(h),
\end{equation}
 plus analogue transformations for the sources, which we do not report here, since they are not important for the purpose of this section, and where $s_1$ and $t_i$ are one-loop coefficients. The effect of~\eqref{eq:fieldred}  is to add terms proportional to the equations of motion, so the non-covariant terms $G_{\text{nc}}(g,h)$ obtained from Feynman diagrams becomes modified as follows
\begin{equation}\label{eq:eomtermgrav}
    G_{\text{nc}}(g,h)\rightarrow G_{\text{nc}}(g,h)-\int\frac{\delta S_{\text{QG}}}{\delta g_{\mu\nu}}\Delta g_{\mu\nu}=G(g).
\end{equation}

The coefficients $t_i$ and $s_1$ can be obtained by working on the term $(S,X)$ and noticing that, by power counting and ghost number, its most general form is 
\begin{equation}\label{eq:Xgravform}
    X(\Phi,K)=\int\Delta g_{\mu\nu}\tilde{K}_g^{\mu\nu}+\int\Delta C^{\rho}K^{C}_{\rho}, \qquad \tilde{K}_g^{\mu\nu}\equiv K_{g}^{\mu\nu}+(\zeta+\square/\alpha)\int\frac{\delta\mathcal{G}_{\rho}}{\delta g_{\mu\nu}}\bar{C}^{\rho}.
\end{equation}
From this expression it is possible to work out the renormalization of the BRST transformations, from which the coefficients in~\eqref{eq:deltagdeltac} are extracted. They read~\cite{Anselmi:2018ibi}
\begin{equation}
    s_1\pi^2\varepsilon=\frac{\alpha  \lambda }{64 \omega ^2}
\end{equation}
\begin{equation}
\begin{split}
    t_1\pi^2\varepsilon &=-\frac{5 \alpha}{18}-\frac{\alpha  \lambda
   }{3}+\frac{\xi }{9}-\frac{\alpha  \lambda }{24 \omega
   ^2}+\frac{\xi }{12 \omega ^2}+\frac{5 \alpha }{18
   \omega }+\frac{5 \xi }{36 \omega },\\[2ex]
t_2\pi^2\varepsilon &=\frac{5 \alpha }{72}-\frac{5 \alpha  \lambda
   }{48}-\frac{\xi }{36}-\frac{\alpha  \lambda }{192
   \omega ^2}-\frac{\xi }{48 \omega ^2}-\frac{5 \alpha
   }{72 \omega }-\frac{5 \xi }{144 \omega }.
   \end{split}
\end{equation}
The coefficient $t_0$ is arbitrary, since it multiplies a covariant term, and encodes a surviving gauge dependence of the counterterms. A convenient way of parametrize it is
\begin{equation}
    t_0\pi^2\varepsilon=\frac{3 \alpha  \lambda }{16}+\frac{\alpha 
   \lambda }{64 \omega ^2}-\frac{3 \xi }{64 \omega
   ^2}-\frac{3 \xi }{16 \omega }+\frac{A}{8},
\end{equation}
where $A$ is a gauge-dependent arbitrary coefficient. The details of this procedure can be found in~\cite{Anselmi:2018ibi}\footnote{Note that the formulas of this paper are related to those in~\cite{Anselmi:2018ibi} through the substitutions $\alpha\rightarrow 1/\alpha$, $\xi\rightarrow 1/\xi$.}. However, in~\autoref{subsec:rank1} we repeat some of the steps for the case of rank-1 field. 

Here and in the next sections the beta functions are related with the renormalization constants by means of the relation $\beta_i=16\pi^2\varepsilon\delta Z_i$. The coefficients $\delta Z_i$ in~\eqref{scount} can be worked out by computing the
graviton self energy and the renormalization of the BRST operators. In the case of pure gravity the results are \cite{Avramidi:1985ki, Salvio:2017qkx, Anselmi:2018ibi}
\begin{eqnarray}
\beta_{\alpha}&=&-\alpha^2\beta_{1/\alpha} =-\frac{133}{10}\alpha^2,\qquad \beta_{\xi}=-\xi^2\beta_{1/\xi} =\frac{5}{6}\xi^2+5\alpha\xi+\frac{5\alpha^{2}}{3},   \notag \\[2ex]
\beta_{\zeta} &=&-\zeta \left( 
\frac{5\xi}{6 }+\frac{5\alpha^2 }{3\xi ^{2}}+A\right), \quad \beta_{\Lambda} =\Lambda\left(5\alpha-2\xi
-2A\right)+\frac{\zeta^{2}}{4}\left(5\alpha^2+\xi^2\right).
\label{betas}
\end{eqnarray}%

Finally, we comment about the signs of the constants $\alpha$ and $\xi$. The theory~\eqref{eq:lhd} propagates a massless graviton, a massive scalar and a massive spin-2 particle. The latter is a ghost and therefore is responsible for the violation of unitarity. As mentioned in the introduction, this problem can be solved by turning the ghost into a purely virtual particle, that is to say particle that can only mediate interactions without ever appear as an on-shell external state. Purely virtual particles are introduced by means of a different quantization procedure called \emph{fakeon prescription}~\cite{Anselmi:2017yux, Anselmi:2017ygm}. In this way, we can consistently project the total Fock space of the theory onto a subspace where the would-be ghost cannot be an external line in the Feynman diagrams. This projection is nontrivial, since it cannot be naively done by just excluding some diagrams. The reason is that for a projection to be consistent it has to be supplemented by some additional procedures, otherwise the degrees of freedom we want to remove would be generated back by loop corrections. One example of a consistent projection is the one done in gauge theories, where the BRST symmetry allows to remove the Faddeev-Popov ghosts from the spectrum, as well as the longitudinal and temporal component of the gauge fields. In the case of purely virtual particles this goal is achieved by a prescription for the scattering amplitudes instead of a symmetry. However, this procedure cannot be applied to tachyons. Therefore, we impose $\alpha>0$ and $\xi>0$ since they are related to the squared masses $\zeta\alpha$ and $\zeta\xi$ of the spin-2 fakeon and the scalar field, respectively. The same rationale is applied to the scalar ghost in~\eqref{eq:rank1flatF}.

\sect{Coupling to Quantum Gravity}
\label{sec:couplingQG}
In this section we couple the theory~\eqref{eq:rank1FlatFInt} to the quantum gravity theory~\eqref{eq:lhd}. The coupling to matter generates contributions to the beta functions~\eqref{betas}, as well as contributions to the renormalization of the field $A_{\mu}$ and to the parameters $m$, $\lambda_2$ and $\lambda_4$. As we show below, the nonminimal couplings $RA_{\mu}A^{\mu}$ and $R^{\mu\nu}A_{\mu}A_{\nu}$ are also generated at one loop. Therefore we add them to classical action. We compare some of the results with Proca theory coupled to quantum gravity, which is nonrenormalizable, as we explicitly show.

Finally, we include the renormalization of a scalar field nonminmally coupled to quantum gravity.
\subsection{Rank-1 Fields}\label{subsec:rank1}
We consider the covariantized version of~\eqref{eq:rank1FlatFInt} and include all possible terms with dimension four\footnote{A $\mathbb{Z}_2$ symmetry prevents to generate terms with odd powers of $A_{\mu}$.}
\begin{figure}[t]
\centering
\includegraphics[width=10cm]{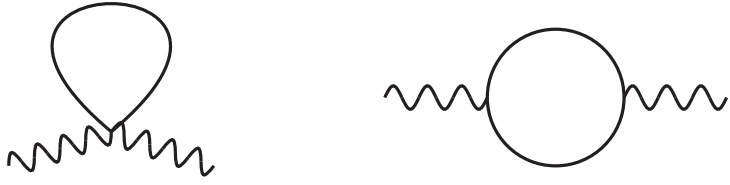}
\caption{Contributions to the graviton self energy due to the presence of matter. The wavy lines denote the graviton, while the solid lines are matter (either scalar or vector fields).}
\label{fig:selfgrav}
\end{figure}
\begin{eqnarray}
\label{eq:rank1FIntGrav}
S_A(A,g)=-\frac{1}{2}\int\sqrt{-g}\left[\frac{1}{2}F_{\mu\nu}F^{\mu\nu}\right.&&\left.+\lambda_2(\nabla_{\mu}A^{\mu})^2-m^2A_{\mu}A^{\mu}\right.\nonumber\\
&&\left.+\eta_1RA_{\mu}A^{\mu}+\eta_2R^{\mu\nu}A_{\mu}A_{\nu}-\frac{\lambda_4}{4}(A_{\mu}A^{\mu})^2\right],
\end{eqnarray}
where $\eta_1$, $\eta_2$ are real parameters. Note that there is no need to add the term $\nabla_{\mu}A_{\nu}\nabla^{\nu}A^{\mu}$ with an independent coupling, since it would be equivalent to a redefinition of $\eta_2$. Then we expand the metric around Minkowski spacetime and compute the relevant diagrams. The contributions to the purely gravitational counterterms~\eqref{scount} are derived from the diagrams in~\autoref{fig:selfgrav}. We find
\begin{equation}
    \frac{\beta_{\alpha}^{A}}{\alpha^2}=-\frac{7}{30}-\frac{\eta _2(1+5\lambda_2)}{6
   \lambda_2}-\frac{\eta _2^2 \left(1+4 \lambda_2+7
   \lambda_2^2\right)}{24 \lambda_2^2},
   \end{equation}
   \begin{equation}\label{eq:deltaxi1}
   \frac{\beta_{\xi}^{A}}{\xi^2}=\frac{1}{6}+3 \eta _1^2 \left(3+\frac{1}{\lambda
   _1^2}\right)-\frac{\eta _1 \left(1-3 \lambda
   _1\right)}{\lambda_2}-\frac{\eta _2 \left(1-7
   \lambda_2\right)}{6 \lambda_2}+\frac{3 \eta _1 \eta
   _2 \left(1+3 \lambda_2^2\right)}{2 \lambda
   _1^2}+\frac{\eta _2^2 \left(5+2 \lambda_2+17 \lambda
   _1^2\right)}{24 \lambda_2^2},
\end{equation}
\begin{equation}
    \beta_{\zeta}^A=m^2\left[\frac{1}{2}-\frac{1}{6 \lambda_2}+\frac{\left(4 \eta
   _1+\eta _2\right) \left(1+3 \lambda_2^2\right)}{4
   \lambda_2^2}\right], \qquad \beta_{\Lambda}^A=-\frac{m^4}{2 \lambda_2^2}(1+3 \lambda_2^2).
\end{equation}
First, we comment on asymptotic freedom in the gravitational sector. The theory~\eqref{eq:lhd} with the no-tachyon conditions $\alpha>0$ and $\xi>0$ is not asymptotically free, as it can be seen from~\eqref{betas}. Moreover, coupling~\eqref{eq:lhd} to standard matter does not change the situation~\cite{Fradkin:1981iu}. However, in~\cite{Piva:2021nyj} it has been shown that including arbitrary-spin generalizations of the theory~\eqref{eq:rank1flatF} can change the sign of $\beta_{\xi}$ without changing that of $\beta_{\alpha}$. Among several solutions, a single rank-3 tensor field theory is enough to achieve asymptotic freedom in the gravitational couplings. Other solutions include a certain number of rank-2 and rank-1 fields, but no solution with rank-1 fields only or fermionic multiplets were found. However, in~\cite{Piva:2021nyj} nonminimal terms, as well as the analogue of $\lambda_2$ for spin larger than 2, were assumed to be small. The calculations of this section are the generalization of those in~\cite{Piva:2021nyj} for the case of rank-1 field and they include all the renormalizable couplings. 

From~\eqref{eq:deltaxi1} we find that $\beta_{\xi}+\beta_{\xi}^A$ cannot be negative, which is a necessary condition for asymptotic freedom. In conclusion, no massive vector fields of any type can change the asymptotic behavior of $\xi$.   

Now we proceed with the renormalization of the terms that contain the field $A_{\mu}$ by using again the Batalin-Vilkovisky formalism. As explained in~\autoref{sec:BV}, renormalziation is a combinations of field and parameters redefinitions. We perform field redefinitions first, so from what is left of the counterterms we can read the renormalization constants of the physical parameters. We introduce the source $K_{A}^{\mu}$ for the BRST transformation of $A_{\mu}$, through the term
\begin{equation}
    S_{K_A}=-\int\mathcal{R}^{A}_{\mu}K_A^{\mu},\qquad \mathcal{R}^A_{\mu}=-C^{\nu}\partial_{\nu}A_{\mu}-A_{\nu}\partial_{\mu}C^{\nu},
\end{equation}
so the extended action now reads
\begin{equation}
    S=S_{\text{QG}}+S_A+(S_K,\Psi)+S_K+S_{K_A}.
\end{equation}
Taking it into account this extension, the functional $X$ in~\eqref{eq:Xgravform} is modified as
\begin{equation}
    X\rightarrow X+\int\Delta A_{\mu}K_{A}^{\mu}, \qquad \Delta A_{\mu}=a_0A_{\mu}+a_1h_{\mu}^{\nu}A_{\nu}+a_2hA_{\mu}+\mathcal{O}(h^2),
\end{equation}
where the last term is the most general functional we can build that involves the field $A_{\mu}$ and has ghost number -1, and dimension 3 and $a_i$ are real coefficients to be determined.
\begin{figure}[t]
\centering
\includegraphics[width=16.3cm]{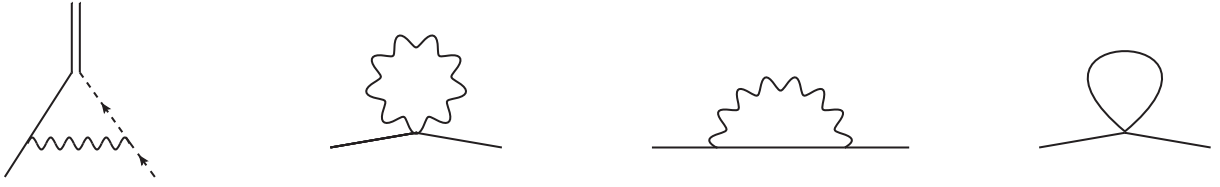}
\caption{The first diagram contributes to the renormalization of the BRST transformation of matter fields (either scalar or vector). The wavy lines denote the graviton, the dashed lines the Faddeev-Popov ghosts, the solid lines the matter fields and the double solid line the sources $K_{A,\phi}$. The other diagrams are the contributions to the matter self energy.}
\label{fig:source}
\end{figure}
In order to obtain $a_i$ we focus on the terms in $(S,X)$ that contain $A_{\mu}$, which read
\begin{equation}\label{eq:sigmaXA}
\begin{split}
    (S,X)\left.\right|_{A}=&\int\frac{\delta S_A}{\delta g_{\mu\nu}}\Delta g_{\mu\nu}+\int\frac{\delta S_A}{\delta A_{\mu}}\Delta A_{\mu}+\int\frac{\delta S_{K_A}}{\delta A_{\mu}}\Delta A_{\mu}+\int\frac{\delta S_{K_A}}{\delta C^{\rho}}\Delta C^{\rho}\\[2ex]
    & +\int\mathcal{R}^{A}_{\mu}\left(\int\frac{\delta \Delta A_{\rho}}{\delta A_{\mu}}K_{A}^{\rho}\right)+\int\mathcal{R}_{\mu\nu}\left(\int\frac{\delta \Delta A_{\rho}}{\delta h_{\mu\nu}}K_{A}^{\rho}\right).
\end{split}
\end{equation}
In particular, the coefficients $a_1$ and $a_2$ can be derived from the renormalization of the BRST transformations by computing the first diagram in~\autoref{fig:source}. 
Therefore, it is enough to look at the terms in~\eqref{eq:sigmaXA} that are linear in $C$ and $K_A$. A straightforward calculation gives
\begin{equation}
    (S,X)\left.\right|_{ACK_A}=\int\left[s_1C^{\nu}\partial_{\nu} A_{\mu}K_A^{\mu}+(s_1-a_1)C^{\nu}A_{\nu}\partial_{\mu}K_A^{\mu}-a_1\partial^{\mu}C_{\nu}A_{\mu}K_A^{\nu}-2a_2 \partial_{\mu}C^{\mu}A_{\rho}K_A^{\rho}\right],
\end{equation}
which can be used to read the coefficients from the first diagram in~\autoref{fig:source}. Finally, the coefficient $a_0$ can be derived by using the second term in~\eqref{eq:sigmaXA} to renormalize the field $A_{\mu}$ in the second and third diagrams in~\autoref{fig:source}. After performing these computations the results are
\begin{equation}
\begin{split}
a_0=\frac{1}{48\pi^2\varepsilon}\left[-19 \xi  \eta _1-\frac{27 \xi  \eta _1}{2 \omega
   }+\frac{10 \alpha  \eta _2}{3}-\frac{23 \xi  \eta
   _2}{6}\right.&\left.-\frac{9 \xi  \eta _2}{4 \omega }+\xi  \eta _1
   \eta _2+\frac{5 \alpha  \eta _2^2}{12}\right.\\
   &\left.+\frac{7 \xi 
   \eta _2^2}{12}-\frac{2 \xi  \eta _1 \eta _2}{\lambda
   _1}-\frac{5 \alpha  \eta _2^2}{6 \lambda_2}-\frac{11
   \xi  \eta _2^2}{12 \lambda_2}\right],
   \end{split}
\end{equation}
\begin{equation}
a_1=\frac{1}{72\pi^2\omega\varepsilon}\left[6\xi\eta_1+(\xi-10\alpha)\eta_2\right],\qquad a_2=-\frac{1}{4}a_1.
\end{equation}
Note that $a_1$ and $a_2$ are not vanishing when the nonminimal couplings are present, which means that the renormalization of the $A_{\mu}$ field is not multiplicative. This is a parametrization and gauge dependent fact and can be avoided by by using other techniques, such as the background field method. However, keeping $a_1$ and $a_2$ nonzero allows to have a better control over the computations and make cross checks of the results.  

The analogue of~\eqref{eq:eomtermgrav} for the noncovariant terms $G_{A,\text{nc}}(h,A)$, obtained from diagrams that contain at least two $A_{\mu}$, reads

\begin{equation}\label{eq:eomtermA}
G_{A,\text{nc}}(h,A)\rightarrow G_{A,\text{nc}}(h,A)-\int\frac{\delta S_{A}}{\delta g_{\mu\nu}}\Delta g_{\mu\nu}-\int\frac{\delta S_{A}}{\delta A_{\mu}}\Delta A_{\mu}=S_{A}^{\text{ct}}(A,g), 
\end{equation}
After this redefinition the counterterms are
\begin{eqnarray}\label{eq:SAct}
    S_{A}^{\text{ct}}(A,g)=-\frac{1}{2}\int\sqrt{-g}&&\left[\frac{1}{2}F_{\mu\nu}F^{\mu\nu}+\lambda_2\delta Z_{\lambda_2}(\nabla_{\mu}A^{\mu})^2-m^2\delta Z_{m^2}A_{\mu}A^{\mu}\right.\nonumber\\
    &&\left.+\eta_1\delta Z_{\eta_1}RA_{\mu}A^{\mu}+\eta_2\delta Z_{\eta_2}R^{\mu\nu}A_{\mu}A_{\nu}-\frac{\lambda_4\delta Z_{\lambda_4}}{4}(A_{\mu}A^{\mu})^2\right],
\end{eqnarray}
where each parameter is renormalized by means of the redefinition of the form
\begin{equation}
    \lambda\rightarrow (1+\delta Z_{\lambda})\lambda,
\end{equation}
where $\lambda$ indicates a generic parameter.
\begin{figure}[t]
\centering
\includegraphics[width=16cm]{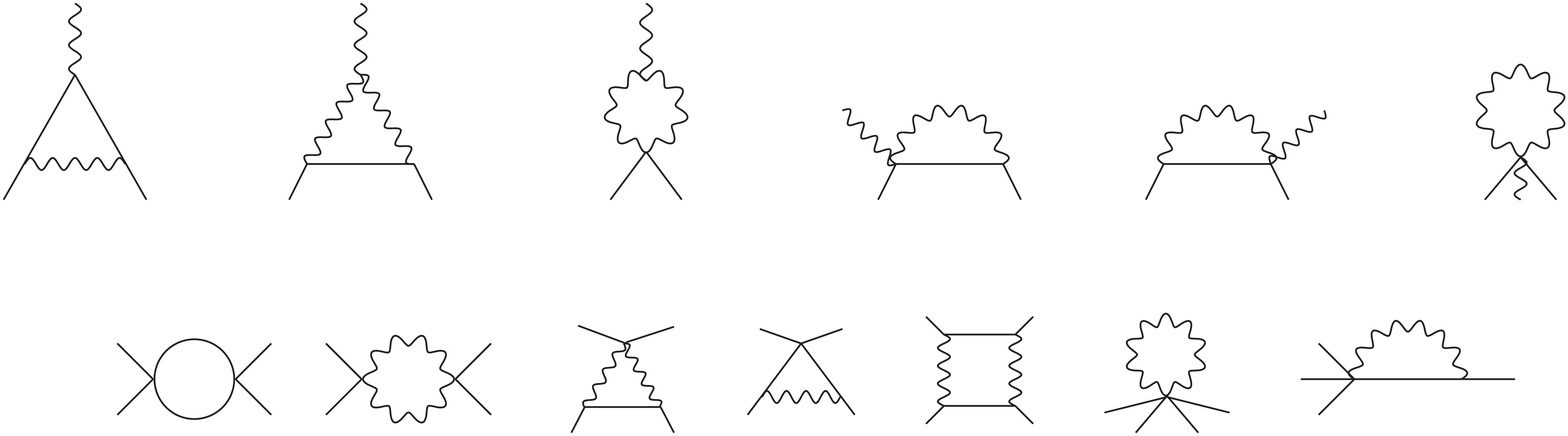}
\caption{In the first line the contributions to the matter-matter-graviton vertex. In the second line the correction to the matter quartic self interactions. In the second line the diagrams obtained from all the inequivalent permutations of external legs have not been depicted for space reasons. They are additional 22 diagrams.}
\label{fig:tribo}
\end{figure}

The renormalization of $m^2$ and $\lambda_2$ can be extracted from the sum of the second, third and fourth diagram in~\autoref{fig:source}. We find

\begin{eqnarray}\label{eq:deltaZm2}
\beta_{m^2}=&&m^2\left[-3\xi-\frac{20 \xi  \eta _1}{3}+9 \xi  \eta _1^2-\frac{70 \alpha  \eta _2}{9}-\frac{14 \xi  \eta
   _2}{9}+\frac{11}{3} \xi  \eta _1 \eta _2-\frac{20 \alpha  \eta _2^2}{9}+\frac{23 \xi  \eta _2^2}{36}\right.\nonumber\\
   &&\left.-\frac{\eta_2}{3\lambda_2}
   \left(4 \xi  \eta _1 +\frac{5\alpha\eta_2}{3}+\frac{11 \xi \eta_2}{6}\right)+\frac{3\xi}{4\lambda_2^2}
   \left(2\eta _1+\eta _2\right)^2-\frac{3\lambda_4(3+\lambda_2^2)}{2\lambda_2^2}-2A\right]\nonumber \\
   &&\left.+\zeta\left[-10 \alpha ^2
   \eta _1+4 \xi ^2 \eta _1+9 \xi ^2 \eta _1^2-\frac{15 \alpha ^2 \eta _2}{2}+\frac{3 \xi ^2 \eta _2}{2}+3
   \xi ^2 \eta _1 \eta _2-\frac{5}{2} \alpha ^2 \eta _2^2+\frac{1}{4} \xi ^2 \eta _2^2\right.\right. \nonumber \\
   &&\left.+\frac{3\xi^2}{4\lambda_2} \left(2\eta_1+\eta_2\right)^2\right].
\end{eqnarray}

\begin{eqnarray}\label{eq:deltaZlambda1}
  \beta_{\lambda_2}= &&-2\lambda_2 \left(5 \alpha +\xi +\frac{10 \xi  \eta _1}{3}+\frac{35 \alpha  \eta _2}{9}+\frac{7 \xi  \eta _2}{9}-\frac{1}{3} \xi 
   \eta _1 \eta _2-\frac{5 \alpha  \eta _2^2}{36}-\frac{7 \xi  \eta _2^2}{36}\right)\nonumber\\
   &&+4 \xi  \eta _1-\frac{10 \alpha  \eta _2}{3}-\frac{2 \xi  \eta _2}{3}-\frac{16}{3} \xi  \eta _1 \eta _2-\frac{20 \alpha  \eta
   _2^2}{9}-\frac{13 \xi  \eta _2^2}{9}+\frac{\eta _2 }{\lambda_2}\left(2 \xi  \eta _1+\frac{5 \alpha  \eta _2}{6}+\frac{\xi  \eta _2}{6}\right).\nonumber\\
   &&
\end{eqnarray}
As expected, $\beta_{\lambda_2}$ is gauge independent, while $\beta_{m^2}$ is not. However, in our parametrization, all the gauge dependence is encoded in the quantity $A$ in the second line of~\eqref{eq:deltaZm2}.

From the diagrams in the first row of~\autoref{fig:tribo} we derive the counterterms of the nonminmal couplings. The results are a few lines long and we write them by collecting the powers of the couplings $\eta_1$, $\eta_2$ and $\lambda_4$. They read
\begin{eqnarray}
    \beta_{\eta_1}=&&\frac{\lambda _4\left(\lambda_2-1\right) }{3 \lambda_2}+2\eta_1\left(5 \alpha +\frac{5 \alpha ^2}{3 \xi }+\xi\right)+5 \eta _2 \left(\frac{2 \alpha }{3}+\frac{\alpha ^2}{2 \xi }+\frac{\xi }{6}\right)\nonumber\\
    &&  +\frac{3\eta _1 \lambda _4}{2 \lambda_2^2}\left(1+3 \lambda_2^2\right)+\frac{ \eta _2 \lambda _4}{6 \lambda_2^2}\left(2-\lambda_2+5 \lambda_2^2\right)+\frac{2 \eta _1 \eta _2}{9} \left(35 \alpha +10 \xi +\frac{3 \xi }{\lambda_2}\right)\nonumber\\
    &&+\frac{14\eta_1^2\xi}{3}+\frac{\eta_2^2}{6} \left(10 \alpha +\frac{5 \alpha ^2}{\xi }+\xi +\frac{2 \xi }{\lambda_2}\right)-\frac{\eta _1^2 \eta _2 \xi  }{6 \lambda_2^2}\left(21-14 \lambda_2+25 \lambda_2^2\right)\nonumber\\
    &&+\frac{\eta_1\eta_2^2}{36}\left(80 \alpha -29 \xi -\frac{45 \xi }{\lambda_2^2}+\frac{20 \alpha }{\lambda_2}+\frac{46 \xi }{\lambda_2}\right)-\frac{3 \eta_1^3 \xi}{\lambda_2^2}\left(1+3 \lambda_2^2\right)\nonumber\\
    &&-\frac{\eta_2^3}{72}\left(\xi -40 \alpha +\frac{9 \xi }{\lambda_2^2}-\frac{6 \xi }{\lambda_2}\right)
\end{eqnarray}

\begin{eqnarray}
    \beta_{\eta_2}=&&4 \lambda_2 (5 \alpha +\xi )+\frac{\lambda _4\left(5 \lambda_2+1\right) }{3 \lambda_2}+\frac{8\eta_1}{3} \xi  \left(5 \lambda_2-3\right)\nonumber\\
    &&+\frac{2 \eta _2}{9}  \left(40 \alpha +5 \xi +70 \alpha  \lambda_2+14 \xi  \lambda_2\right)+\frac{\lambda_4\eta_2}{6} \left(7+\frac{4}{\lambda_2}+\frac{1}{\lambda_2^2}\right)\nonumber\\
    &&+\frac{2\eta_1\eta_2}{3} \xi  \left(5-\frac{9}{\lambda_2}-2 \lambda_2\right)-2 \eta _1^2 \xi  \left(5+\frac{1}{\lambda_2}\right)\nonumber\\
    &&+\frac{\eta_2^2}{90}\left(90 \alpha +\frac{39 \xi }{2}-\frac{5 \alpha }{\lambda_2}-\frac{17 \xi }{2 \lambda_2}-5 \alpha  \lambda_2-7 \xi  \lambda_2\right)-\frac{\eta_1^2\eta_2\xi}{\lambda_2^2}\left(1+4 \lambda_2+7 \lambda_2^2\right)\nonumber\\
    &&-\frac{\eta_1\eta_2^2\xi}{3 \lambda_2^2}\left(3+4 \lambda_2+9 \lambda_2^2\right)+\frac{\eta_2^3}{36}\left(-21 \xi -\frac{9 \xi }{\lambda_2^2}+\frac{20 \alpha }{\lambda_2}+\frac{10 \xi }{\lambda_2}\right).
\end{eqnarray}

Finally, from the diagrams in the second row of~\autoref{fig:tribo}, plus those with inequivalent exchanges of external legs, we derive the counterterm of the self interaction obtaining

\begin{eqnarray}
    \beta_{\lambda_4}=&&\frac{4\lambda _4}{3}(\xi -10 \alpha )-\frac{\lambda _4^2 }{2 \lambda
   _1^2}\left(5+2 \lambda_2+17 \lambda_2^2\right)+\frac{8\lambda _4  \eta _1}{3 \lambda_2}\xi\left(3+\lambda_2\right)\nonumber\\
   &&-\frac{4\lambda _4\eta _2}{9} \left(50 \alpha +\xi -\frac{9 \xi }{\lambda_2}\right)+\frac{4 \lambda _4 \eta _1 \eta _2 }{3 \lambda_2^2}\xi  \left(9+4 \lambda_2+13 \lambda_2^2\right)+\frac{12 \lambda _4 \eta _1^2}{\lambda_2^2}\xi  \left(1+\lambda_2+4 \lambda_2^2\right)\nonumber\\
   &&-\frac{\lambda _4 \eta _2^2}{9} \left(70 \alpha -19 \xi -\frac{27 \xi }{\lambda_2^2}+\frac{10 \alpha }{\lambda_2}+\frac{2 \xi }{\lambda_2}\right)-20 \eta _1^2 \left(\alpha ^2-\alpha  \xi +\xi ^2\right)\nonumber\\
   &&-\frac{\eta _2^2}{18} \left(275 \alpha ^2-40 \alpha  \xi +53 \xi ^2\right)+2 \eta _1^2 \eta _2 \xi  \left(20 \alpha -21 \xi -\frac{25 \xi }{\lambda_2}\right)\nonumber\\
   &&-\eta _1 \eta _2^2 \left(10 \alpha ^2-20 \alpha  \xi +9 \xi ^2+\frac{23 \xi ^2}{\lambda_2}\right)-\frac{12 \eta _1^3 \xi ^2 }{\lambda_2}\left(3+5 \lambda_2\right)\nonumber\\
   &&+\frac{\eta_2^3}{18} \left(-200 \alpha ^2+40 \alpha  \xi -11 \xi ^2-\frac{63 \xi ^2}{\lambda_2}\right)-\frac{12 \eta _1^3 \eta _2 \xi ^2 }{\lambda_2^2}\left(3+4 \lambda_2+5 \lambda_2^2\right)\nonumber\\
   &&+\eta _1 \eta _2^3\xi \left(\frac{20 \alpha }{3}-\frac{5 \xi}{3}-\frac{9 \xi}{\lambda_2^2}-\frac{4 \xi}{\lambda
   _1}\right)+\eta_1^2\eta_2^2\xi \left(20 \alpha -15 \xi -\frac{27 \xi }{\lambda_2^2}-\frac{22 \xi }{\lambda_2}\right)\nonumber\\
   &&-\frac{18 \eta _1^4 \xi ^2 }{\lambda_2^2}\left(1+2 \lambda_2+5 \lambda_2^2\right)-\frac{\eta _2^4}{4}\left(\frac{85 \alpha ^2}{9}-\frac{20 \alpha  \xi }{9}+\frac{5 \xi ^2}{18}+\frac{9 \xi ^2}{2
   \lambda_2^2}+\frac{\xi ^2}{\lambda_2}\right)
\end{eqnarray}
We report the gauge-independent beta functions in some specific cases, in order to show that all the terms in~\eqref{eq:rank1FlatFInt} must be included if $\lambda_2\neq 0$. First, when the nonminmal couplings are vanishing the beta functions become
\begin{equation}\label{eq:beta0lambda}
    \beta_{\lambda_2}=-2 (5 \alpha +\xi ) \lambda_2,\qquad \beta_{\lambda_4}=\frac{4\lambda _4}{3}(\xi -10 \alpha )-\frac{\lambda _4^2 }{2 \lambda
   _1^2}\left(5+2 \lambda_2+17 \lambda_2^2\right),
\end{equation}
\begin{equation}\label{eq:beta0eta}
    \beta_{\eta_1}=\frac{\lambda _4 \left(\lambda_2-1\right)}{3 \lambda_2},\qquad \beta_{\eta_2}=4 \lambda_2 (5 \alpha +\xi )+\frac{\lambda _4\left(5 \lambda_2+1\right) }{3 \lambda_2}.
\end{equation}
This tells us that the nonminimal terms are generated anyway at one loop, even if they are absent at tree level. If we also set $\lambda_4$ to zero, $\beta_{\eta_1}$ vanishes but $\beta_{\eta_2}$ does not. Then, we could choose 
\begin{equation}\label{eq:conditionbeta}
5\alpha+\xi=0
\end{equation}
and all the beta functions in~\eqref{eq:beta0lambda} and~\eqref{eq:beta0eta} would be zero. However, the condition~\eqref{eq:conditionbeta} is not renormalization-group invariant and violates one of the no-tachyon conditions $\alpha>0$, $\xi>0$. Therefore, it is not a viable choice. Barring this possibility, this means that the presence of the term $(\nabla^{\mu}A_{\mu})^2$ in the free action forces us to have $\eta_2\neq 0$, which in turn makes $\beta_{\eta_1}$ and $\beta_{\lambda_4}$ nonzero. In conclusion, all the terms in~\eqref{eq:rank1FIntGrav} must be included.

\subsection{Proca fields}\label{subsec:proca}
We present the one-loop counterterms in the case of Proca theory in order to make a comparison with the results obtained in the previous subsection. 

First we consider the free Proca action
\begin{equation}
S_{\text{P}}(A,g)=-\frac{1}{2}\int\sqrt{-g}\left(\frac{1}{2}F_{\mu\nu}F^{\mu\nu}-m^2A_{\mu}A^{\mu}\right)
\end{equation}
minimally coupled to quantum gravity~\eqref{eq:lhd}. In this case, the theory is renormalizable. The reason is that $F_{\mu\nu}$ does not contain covariant derivatives and an external $A_{\mu}$ line is necessarily multiplied by either a mass or a derivative (pretty much like in the case of the free scalar described below). Therefore, any diagram with an external $A_{\mu}$ leg can only renormalize either the kinetic or mass terms. Higher-derivative terms cannot be generated because the counterterms are polynomial in dimensionful parameters, so no inverse powers of $m$ can appear to compensate the higher-dimensional operators. This changes if we include self interactions, as shown below. 

The gravitational counterterms are well known~\cite{Fradkin:1981iu} and read
\begin{equation}\label{eq:selfgravctproca}
    S_{\text{grav}}^{\text{ct}}(g)=-\frac{1}{(4\pi)^2\varepsilon}\int\sqrt{-g}\left[3m^4-\frac{m^2}{2}R+\frac{1}{72}R^2+\frac{13}{120}C^2\right],
\end{equation}
while for the other counterterms we find
\begin{equation}
    \delta Z_A=\delta Z_{\lambda_4}=\delta Z_{\eta_1}=\delta Z_{\eta_2}=0, \qquad \delta Z_{m^2}=-\frac{1}{8\pi^2\varepsilon}\left(\frac{3 \xi}{2}+A\right),
\end{equation}
which shows that the nonminimal couplings, as well as the quartic self interaction, are not generated by renormalization if they are absent in the classical action, as explained above. Moreover, the field renormalization is zero, as in the massless case. Note that this is true even with a nonzero cosmological constant, differently from the case of the Einstein-Hilbert action, where the counterterms are polynomials in $\Lambda$ and $1/\zeta$, so it is possible to generate $F_{\mu\nu}F^{\mu\nu}$ at one loop~\cite{Toms:2009vd}.

If we turn on a quartic interaction term $(A_{\mu}A^{\mu})^2$ we find that the divergent part of two-point function still renormalizes the mass term without introducing any higher-dimensional operators
\begin{equation}
    \delta Z_{m^2}=-\frac{1}{8\pi^2\varepsilon}\left(\frac{3 \xi}{2}+A+\frac{9}{4}\lambda_4\right).
\end{equation}
However, the divergent part of the four-point function produces all possible operators with up to four derivatives such as
\begin{equation}
    A_{\mu}A_{\nu}\partial^{\mu}\partial^{\nu}(A_{\rho}A^{\rho}), \qquad A_{\mu}A^{\mu}\square(A_{\nu}A^{\nu}), \qquad A_{\mu}A_{\nu}\partial^{\mu}\partial^{\nu}\partial^{\rho}\partial^{\sigma}(A_{\rho}A_{\sigma})
\end{equation}
and so on. This is true already in flat spacetime. In fact, neglecting the contributions from gravity, we find
\begin{eqnarray}
    &&\langle A_{\mu}(p_1)A_{\nu}(p_2)A_{\rho}(p_3)A_{\sigma}(p_4)\rangle_{1,\text{div}}=\frac{\lambda_4^2}{240\pi^2m^4\varepsilon}\left[2p_{\mu}p_{\nu}p_{\rho}p_{\sigma}+\frac{7p^2}{2}\left(p_{\mu}p_{\nu}\eta_{\rho\sigma}+p_{\rho}p_{\sigma}\eta_{\mu\nu}\right)\right. \nonumber \\
    &&\left.+\frac{p^2}{4}\left(p_{\mu}p_{\rho}\eta_{\nu\sigma}+p_{\mu}p_{\sigma}\eta_{\nu\rho}+p_{\nu}p_{\rho}\eta_{\mu\sigma}+p_{\nu}p_{\sigma}\eta_{\mu\rho}\right)+\frac{(p^2)^2}{4}\left(\eta_{\mu\rho}\eta_{\nu\sigma}+\eta_{\mu\sigma}\eta_{\nu\rho}\right)+\frac{13(p^2)^2}{2}\eta_{\mu\nu}\eta_{\rho\sigma}\right]\nonumber \\
    && +\frac{\lambda_4^2}{32\pi^2m^2\varepsilon}\left[-4\left(p_{\mu}p_{\nu}\eta_{\rho\sigma}+p_{\rho}p_{\sigma}\eta_{\mu\nu}\right)-\left(p_{\mu}p_{\rho}\eta_{\nu\sigma}+p_{\mu}p_{\sigma}\eta_{\nu\rho}+p_{\nu}p_{\rho}\eta_{\mu\sigma}+p_{\nu}p_{\sigma}\eta_{\mu\rho}\right)\right.\nonumber \\
    &&\left.+\frac{p^2}{3}\left(\eta_{\mu\rho}\eta_{\nu\sigma}+\eta_{\mu\sigma}\eta_{\nu\rho}\right)-\frac{14p^2}{3}\eta_{\mu\nu}\eta_{\rho\sigma} \right]+(\nu\leftrightarrow \rho,p\rightarrow k)+(\mu\leftrightarrow\rho, p\rightarrow q)\nonumber \\
    &&-\frac{15\lambda_4^2}{32\pi^2\varepsilon}(\eta_{\mu\nu}\eta_{\rho\sigma}+\eta_{\mu\rho}\eta_{\nu\sigma}+\eta_{\mu\sigma}\eta_{\nu\rho}),
\end{eqnarray}
where
\begin{equation}
    p\equiv p_1+p_2, \qquad k\equiv p_1+p_3, \qquad q\equiv p_2+p_3.
\end{equation}
On the other hand, if we include the nonminimal interactions we find nonrenormalizable operators also in the two-point function. In that case we find
\begin{equation}\label{eq:nmnrterms}
    \langle A_{\mu}(p)A_{\nu}(-p)\rangle_{\text{1,div}}=\frac{\eta_2p^2}{576 \pi^2m^2\varepsilon}\left\{\left[12 \xi\eta_1+(5\alpha+7\xi)\eta_2\right]p^2\eta_{\mu\nu}+2\left[12 \xi\eta_1+(5\alpha-2\xi)\eta_2\right]p_{\mu}p_{\nu}\right\},
\end{equation}
which generates the terms
\begin{equation}
    F_{\mu\nu}\square F^{\mu\nu}, \qquad \partial^{\mu}A_{\mu}\square\partial^{\nu}A_{\nu}.
\end{equation}
In~\eqref{eq:nmnrterms} we have written only the terms that contribute to higher-dimensional operator proportional to the gravitational couplings. Besides them, there are also contributions to the mass and wave-function renormalizations. In particular, since inverse powers of $m^2$ are present, the cosmological constant contributes to both. 

These results are expected since the nonminmal terms spoil renormalizability, as discussed in~\cite{Toms:2015fja}.

To summarize, free Proca theory minimally coupled to quantum gravity is renormalizable as long as also the gravitational theory is renormalizable. On the other hand, adding self interactions and/or nonminmal couplings to gravity spoils renormalizability

\subsection{Scalars}\label{subsec:scalar}
For completeness, we show the results in the case of a scalar field nonminimally coupled to~\eqref{eq:lhd}. The diagrammatic and some details are similar to the rank-1 field case.

The action reads
\begin{equation}\label{eq:scalarNM}
S_{\phi}(\phi,g)=\frac{1}{2}\int\sqrt{-g}\left(\partial_{\mu}\phi\partial_{\nu}\phi g^{\mu\nu}-m^2\phi^2+\eta R\phi^2-\frac{\lambda_4}{12}\phi^4\right), 
\end{equation}
where $\eta$ and $\lambda_4$ are real parameters. Expanding the metric around flat spacetime, we derive the necessary graviton-scalar vertices and compute the diagrams, which are the same as in~\autoref{fig:selfgrav},~\autoref{fig:source}, and~\autoref{fig:tribo} with the solid lines representing the scalar field instead of the massive rank-1 field. From the graviton self-energy diagrams of~\autoref{fig:selfgrav} we derive the contributions of the scalar to the gravitational couplings. We find the known result for the pure gravitational counterterms
\begin{equation}\label{eq:selfgravctscalar}
    S_{\text{grav}}^{\text{ct}}(g)=-\frac{1}{(4\pi)^2\varepsilon}\int\sqrt{-g}\left[\frac{m^4}{2}+\frac{m^2}{6}(1-6\eta)R+\frac{1}{72}(1-6\eta)^2R^2+\frac{1}{120}C^2\right].
\end{equation}
Then we follow the steps of~\autoref{subsec:rank1} and introduce the term
\begin{equation}
S_{K_{\phi}}=-\int C^{\mu}\partial_{\mu}\phi K_{\phi},
\end{equation}
that accounts for the composite BRST operator of $\phi$ and its source $K_{\phi}$. The extended action~\eqref{eq:extact} is
\begin{equation}\label{eq:extactphi}
    S=S_{\text{QG}}+S_{\phi}+(S_K,\Psi)+S_K+S_{K_{\phi}}.
\end{equation}
In analogy with the massive vector case, we perform the field redefinition that turns the noncovariant term $G_{\text{nc}}(h,\phi)$ into
\begin{equation}\label{eq:eomtermphi}
G_{\phi,\text{nc}}(h,\phi)\rightarrow G_{\phi,\text{nc}}(h,\phi)-\int\frac{\delta S_{\phi}}{\delta g_{\mu\nu}}\Delta g_{\mu\nu}-a_0\int\frac{\delta S_{\phi}}{\delta \phi}\phi=S_{\phi}^{\text{ct}}(\phi,g), 
\end{equation}
where $\Delta g_{\mu\nu}$ is given by~\eqref{eq:deltagdeltac} and
\begin{equation}
a_0=-\frac{1}{8\pi^2\varepsilon}\left[\frac{3 \xi}{4}-\frac{3 \eta \xi}{\omega}(\omega+1)+\frac{A}{2}\right].
\end{equation}

In principle there might be an additional term in the redefinition of $\phi$ proportional to $h^{\rho}_{\ \rho}\phi$. However, this term is multiplied by a vanishing coefficient. This can be explicitly check by computing the divergent part of the first diagram in~\autoref{fig:source}. After this operation we are left with
\begin{equation}
    S_{\phi}^{\text{ct}}(\phi,g)=\frac{1}{2}\int\sqrt{-g}\left[\partial_{\mu}\phi\partial_{\nu}\phi g^{\mu\nu}-m^2\delta Z_{m^2}\phi^2+\eta\delta Z_{\eta}R\phi^2-\frac{1}{12}\lambda_4\delta Z_{\lambda_4}\phi^4\right].
\end{equation}
Since in this case the renormalization of $\phi$ is multiplicative, it is easy to derive the correction to the wave function renormalization $\delta Z_{\phi}$ from the second and third diagram of~\autoref{fig:source} and check that
\begin{equation}
\delta Z_{\phi}=2a_0.
\end{equation}
From the same diagrams, together with the fourth one, we derive the beta function of the mass 
\begin{equation}
\beta_{m^2}=2m^2\left\{\frac{\lambda_4
   }{2}+5 \alpha-\frac{\xi}{2}\left[1+12\eta(1-\eta)\right]-\frac{A}{2}\right\}-2\zeta\eta\left[5\alpha^2+\xi^2(1-6\eta)\right]
\end{equation}

From the diagrams in the first row of~\autoref{fig:tribo} we extract the beta function of the nonminimal coupling, which reads
\begin{equation}\label{eq:dZeta}
    \beta_{\eta}=4\eta\left[-\frac{5 \alpha ^2}{6 \xi }+\frac{\xi }{3}-\frac{\lambda_4  \left(1-6 \eta\right)}{24 \eta}-\frac{\eta\xi}{2}\left(5-6\eta\right)\right].
\end{equation}
Moreover, as a consistency check, we have verified the Ward identities of diffeomorphisms that relate the second and third diagram in~\autoref{fig:source} with those in the first row of~\autoref{fig:tribo}.

Finally, from the second row of diagrams in~\autoref{fig:tribo} we extract the beta function of the coupling $\lambda_4$
\begin{equation}\label{eq:dZlam4S}
    \beta_{\lambda_4}=2\lambda_4\left[\frac{3 \lambda_4}{2}+5 \alpha  \left(1+\frac{6 \alpha  \eta^2}{\lambda_4}\right)+\xi  \left(1-6 \eta\right)^2
   \left(1+\frac{6 \xi  \eta^2}{\lambda_4}\right) \right].
\end{equation}

As expected,~\eqref{eq:dZeta} and~\eqref{eq:dZlam4S} are gauge independent and agree with the results in the literature (e.g.~\cite{Salvio:2014soa}). 

Note that the quartic interaction and the nonminimal coupling turn each other on if one of the two is present. Only if they are both absent they are not generated by renormalization. The reason is the same as in Proca theory: when $\lambda_4=\eta=0$ every external $\phi$ line comes with either a derivative or a mass. Therefore, any one-loop diagram can only correct the kinetic term or the mass term. However, once one of the two interactions is nonzero, this is not true anymore and every term of dimension smaller or equal than four that satisfies diffeomorphism invariance is generated.

\sect{Conclusions}
\label{sec:concl}
We have studied the renormalization of the most general massive vector field theory, where all the four components of the multiplet propagates, coupled to quantum gravity. The theory is renormalizable by power counting, even when self interactions are switched on. In particular, all the nonminimal couplings with gravity are generated by renormalization and therefore must be included in the classical action. We compute all the relevant one-loop diagrams and derive the beta functions for the parameters of the theory. We show that there is no choice for their initial value that avoids the generation of nonminimal terms or quartic self interactions. Moreover, the presence of nonminimal couplings cannot change the ultraviolet behavior of the gravitational interactions, which are not asymptotically free in absence of tachyons. Finally, we have made a comparison with the case of Proca theory, which is not renormalizable when self interactions and/or nonminimal couplings with gravity are included, and explicitly show that higher-dimensional divergent terms are generated.

\subsection*{Acknowledgments}
We are grateful to C. Marzo for useful discussions.

\bibliographystyle{JHEP} 
\bibliography{mybiblio}

\end{document}